\documentclass[twocolumn,prb,showpacs,floats,superscriptaddress]{revtex4}
\usepackage[dvips]{graphicx}
\usepackage{amsmath,bm}
\usepackage{psfrag}

\begin{document}

\title{Linear oscillations of a compressible hemispherical bubble \\ on a solid substrate}

\author{Sergey Shklyaev}
\affiliation{Department of Theoretical Physics, Perm State
University, Bukirev 15, Perm 614990, Russia}

\author{Arthur V. Straube\footnote{E-mail: arthur.straube@gmail.com \\[0.5mm] {\mbox{} Paper published in Phys. Fluids {\bf{20}}, 052102 (2008)}}$^{,}$ }
\affiliation{Department of Physics and Astronomy, University of Potsdam, Karl-Liebknecht-Str. 24/25, D-14476 Potsdam-Golm, Germany}


\begin{abstract}
The linear natural and forced oscillations of a compressible
hemispherical bubble on a solid substrate are under theoretical
consideration. The contact line dynamics is taken into account by
application of the Hocking condition, which eventually leads to
nontrivial interaction of the shape and volume oscillations.
Resonant phenomena, mostly pronounced for the bubble with the
fixed contact line or with the fixed contact angle, are found out.
A double resonance, where independent of the Hocking parameter an
unbounded growth of the amplitude occurs, is detected. The
limiting case of weakly compressible bubble is studied. The
general criterion identifying whether the compressibility of a
bubble can be neglected is obtained.
\end{abstract}




\maketitle


\section{Introduction} \label{sec:intro}

The dynamics of bubbles and droplets is of great interest for
their numerous applications. Bubbly fluids are widely used as
displacing fluids in petroleum industry. Small bubbles and
droplets allow to efficiently control heat and mass transfer in
heat-exchangers and reactors and to intensify mixing in
microdevices.\cite{nigmatulin-91, squires-quake-05} Whereas
oscillations of drops suspended in a fluid ambient away from the
boundaries have been scrutinized for over a century,
\cite{zapryanov-tabakova-99} oscillations of drops and bubbles in
contact with solid surfaces have only received attention for the
last few decades. Understanding fundamental aspects of drops and
bubbles interaction with the solid surface is closely related to
the problem of wetting.\cite{dussan-79, de-gennes-85} This
knowledge is of practical importance because many technological
processes deal with spreading of a liquid (a paint, a lubricant,
or a dye) over solid surfaces. From the theoretical point of view,
the presence of a solid surface often meets another problem, the
contact line dynamics, which is currently far from being fully
understood.

Qualitatively, oscillations of a liquid drop of averaged radius
$R$ can be characterized by three time scales: the viscous relaxation
time $\tau_v=R^2/\nu$, the capillary time scale $\tau_c=\sqrt{\rho
R^3/\sigma}$, which is related to the period of the shape
oscillations,\cite{rayleigh-45} and the time scale of the acoustic
oscillations $\tau_a=R/c$. Here, $\sigma$ is the surface tension,
$\rho$, $\nu$, and $c$ are the the density, kinematic viscosity,
and speed of sound, respectively. Quite often, these three time scales relate to each
other as
\begin{equation} \label{sec:intro-times}
\tau_a \ll \tau_c \ll \tau_v.
\end{equation}
This hierarchy of the times allows for the description of the drop
oscillations within the model of inviscid incompressible fluid,
where the dynamics of the drop corresponds to the shape
oscillations. For instance, these inequalities hold for water
drops of $R > 10^{-5}\,{\rm cm}$. For a liquid drop immersed in
another liquid the hierarchy of the time scales is the same. The
only difference is in the meaning of $\rho$, $\nu$, and $c$, which
should now be considered as some effective quantities, e.g., the
mean values for the two liquids.

For a gaseous bubble in an ambient liquid the situation is similar
to the case of the drop, but an additional, the so-called
``breathing mode,'' appears.\cite{rayleigh-17} This mode
corresponds to the radial (volume) oscillations of the bubble and
is caused by the bubble compressibility. Note that since the
densities of the gas and liquid differ considerably, the gas
compressibility is non-negligible even at subacoustic frequencies,
$\omega_b \tau_a \ll 1$, where $\omega_b$ is the frequency of the
breathing mode. In this case, the gas pressure in the bubble very
quickly adjusts to the instant volume of the bubble. Although the
pressure field in the bubble changes in time, its instant
distribution can be considered as spatially homogeneous. If
dissipation is insignificant during a period of oscillation,
the gas can be described by the adiabatic law.
For this situation\cite{wijngaarden-72} $\omega_b=\sqrt{3\gamma
P_g /\rho R^2}$, where $\gamma$ is the adiabatic exponent and
$P_g$ is the equilibrium pressure in the bubble (for a correction
of $\omega_b$ caused by surface tension see
Sec.~\ref{sec:natural-oscillations}). Detailed analyses of damping effects can be found in
Refs.~\onlinecite{wijngaarden-72, nigmatulin-91}. Particularly,
these studies figure out the inequalities
\begin{equation}\label{sec:intro-nodamping}
\omega_b \tau_v \gg 1, \quad \omega_b \tau_t \gg 1, \quad \omega_b
\tau_a \ll 1,
\end{equation}
\noindent which can be applied to neglect the effects caused by
viscosity, heat transfer, and acoustic irradiation, respectively.
Here, we introduce the characteristic time of thermal diffusion
$\tau_t=R^2/\chi_m$ with $\chi_m$ being the maximal from heat
diffusivities of the fluid $\chi$ and the gas $\chi_g$.
As it follows from (\ref{sec:intro-nodamping}), the predominant
damping mechanism for an air bubble in water is heat dissipation.
This mechanism can be neglected provided that $R>10^{-3}\,{\rm
cm}$.

If the frequencies of the shape and volume oscillations become
comparable, the oscillations of these types start to interact. For
instance, parametric excitation of the shape oscillations on top
of the forced breathing mode has been addressed by Mei and
Zhou,\cite{mei-zhou-91} who have found an instability of the
radial oscillations and performed a weakly nonlinear analysis.
Since then, the problem of the parametric instability has received
much attention, see, e.g., a recent review by Feng and
Leal.\cite{feng-leal-97} Another example of the interaction has
been provided by Longuet-Higgins.\cite{longuet-higgins-part1-89}
He focuses on bubble oscillations in a liquid and shows that the
nonlinear coupling of the shape oscillations can lead to the
excitation of the volume mode. Nonlinear oscillations of an
incompressible drop with the accurate account for the dynamics of
the ambient gas have been studied in
Ref.~\onlinecite{alabuzhev-shklyaev-07}. The nonlinear coupling of
the shape oscillations results in generation of sound. It is worth
noting that in the previous studies the interaction of the shape
and volume oscillations arises as a nonlinear effect. In our paper
we report on another, pure linear, mechanism of coupling of the
oscillations of the different kinds. This mechanism of linear
coupling is caused by the contact line dynamics.

Last years have witnessed growing interest in understanding the
contact line dynamics. Although the steady motion of the contact
line has been well studied,\cite{de-gennes-85, voinov-76} there
has been no rigorous theory for unsteady motion yet. What is
typically applied in this situation is a simplified approach. The
thin viscous boundary layer is neglected and a phenomenological
boundary condition imposed on the apparent contact angle is
applied instead. Such condition has been proposed by Hocking for
small oscillations of the contact line.\cite{hocking-87}  The
velocity of the contact line is assumed to be proportional to the
deviation of the contact angle from its equilibrium value (for
simplicity, the equilibrium contact angle is considered to be
$\pi/2$):
\begin{equation} \label{sec:intro-hocking}
\frac{\partial \zeta}{\partial t}=\Lambda \, {\mathbf
n}\cdot\nabla \zeta.
\end{equation}
\noindent Here, $\zeta$ is the deviation of the interface from its
equilibrium position, ${\mathbf n}$ is the external normal to the
solid surface. The coefficient $\Lambda$ has the dimension of
velocity and is referred to as the wetting or the Hocking
parameter. The mostly studied cases correspond to the simplest
limiting situations of either the fixed contact line ($\zeta=0$,
the pinned-end edge condition) or the fixed contact angle
(${\mathbf n}\cdot\nabla\zeta=0$, the free-end edge condition).
Except for these two particular cases, the Hocking condition
(\ref{sec:intro-hocking}) leads to energy dissipation at the
contact line. A simple generalization of
Eq.~(\ref{sec:intro-hocking}) that is based on experimental
observations\cite{dussan-79} and accounts for hysteresis of the
contact line has been provided in
Ref.~\onlinecite{hocking-hysteresis-87}.

A consideration of natural oscillations of a hemispherical drop
for the fixed contact angle has shown \cite{myshkis-92} that the
eigenfunctions coincide with the even modes of the natural
oscillations of a spherical drop. The problem of axisymmetric
oscillations of a drop with the fixed contact line has been
addressed both experimentally\cite{bisch-etal-82,
de-paoli-etal-92} and numerically.\cite{siekmann-schilling-89,
basaran-de-paoli-94, wilkes-basaran-97, wilkes-basaran-99}
Experimental studies\cite{bisch-etal-82, de-paoli-etal-92} focus
on the forced oscillations. In particular, eigenfrequencies are
determined for different equilibrium contact angles. Numerical
investigations deal with natural oscillations of an inviscid
drop,\cite{siekmann-schilling-89} and with
natural\cite{basaran-de-paoli-94} and
forced\cite{wilkes-basaran-97,wilkes-basaran-99} nonlinear
oscillations of a viscous drop. Longitudinal vibrations have
recently been studied theoretically\cite{dong-etal-06} and
experimentally.\cite{noblin-etal-04} In
Ref.~\onlinecite{dong-etal-06} the horizontal and vertical
orientation of the substrate is considered. For the vertical
orientation, gravity leads to asymmetry of oscillations. The study
by Noblin {\it et al.}\cite{noblin-etal-04} has manifested the
nontrivial dynamics of the contact line: at a relatively small
amplitude of vibration the contact line remains pinned, while at
higher amplitudes it starts to move in the stick-slip regime.
Asymmetric vibrations as well as possible microfluidic applications
have been discussed in Ref.~\onlinecite{daniel-etal-05}.

Recently, the focus of attention has been on the impact of the
contact line dynamics on the natural and forced oscillations of an
{\it incompressible} hemispherical drop on a solid
substrate.\cite{lyubimov-etal-06, lyubimov-etal-04} Axisymmetrical
modes of the natural oscillations caused by transversal vibrations
of the substrate are studied in
Ref.~\onlinecite{lyubimov-etal-06}. Another
paper\cite{lyubimov-etal-04} addresses the nonaxisymmetrical modes
of the natural oscillations and the forced oscillations for the
longitudinal vibrations of the substrate, where inertia of the
ambient fluid is taken into account. To some extent, these
analyses can be applied to describe oscillations of an {\it
incompressible} bubble in a liquid. However, the bubble
compressibility, a principal feature of the present paper, can
become of crucial importance, which has been beyond the scope of
the previous research.\cite{lyubimov-etal-06, lyubimov-etal-04} 
A recent experimental study\cite{zoueshtiagh-etal-06} has
indicated an interesting crossover, where beyond a certain
threshold of vibration acceleration the bubble can split into
smaller parts. Despite noticeable progress, the dynamics of a {\it
compressible} bubble on a vibrated substrate has not been fully
understood.

In the present paper we address the behavior of a compressible
hemispherical bubble on a solid substrate. The paper is outlined
as follows. In Sec.~\ref{sec:natural-oscillations} we analyze
natural oscillations. Section~\ref{sec:forced-oscillations} deals
with the forced oscillations for the normally vibrated substrate.
A transition to the case of a weakly compressible bubble is
performed in Sec.~\ref{sec:weak-compressibility}. This is the
situation, when the frequency of the volume oscillations is high
compared with that for the shape oscillations. Particularly, we
obtain a general criterion identifying whether compressibility of
a bubble can be neglected. In Sec.~\ref{sec:conclusions} we
discuss the results and summarize the most important conclusions.

\section{Natural oscillations} \label{sec:natural-oscillations}

%
\begin{figure}[!t]
\centering
\includegraphics[width=0.28\textwidth]{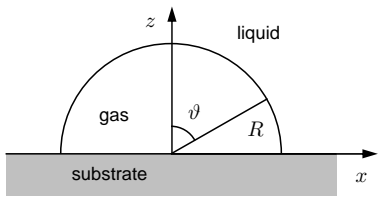}
\caption{The geometry of the natural oscillations problem.} \label{fig1}
\end{figure}
%

Consider natural oscillations of a compressible gaseous bubble
that sits on a solid substrate and is surrounded by a liquid,
Fig.~\ref{fig1}. Suppose that in equilibrium the bubble is
hemispherical, i.e., the equilibrium contact angle equals $\pi/2$
(although this assumption is of no crucial importance, it
considerably simplifies the forthcoming analysis). We assume that
the gas density is small enough so that the frequency of the
volume oscillations is comparable with that of the shape
oscillations. Next we impose the frequency restrictions
(\ref{sec:intro-nodamping}), which ensure insignificance of the
damping effects caused by acoustic irradiation, viscous and heat
dissipation. Finally, we admit that the bubble is sufficiently
small so that the hydrostatic difference of the pressure is
negligible compared with the pressure contribution caused by
surface tension. This assumption allows us to neglect gravity and
ensures that the averaged bubble surface is hemispherical to very
high accuracy. Formally, this situation implies that the Bond
number, ${\rm Bo}=\rho g R^2/\sigma$, is small. Here $g$ is the
acceleration due to gravity. For instance, even under terrestrial
conditions, an air bubble of radius $R=0.1\;{\rm cm}$ suspended in
water is characterized by ${\rm Bo}\approx 0.1$. Thus, the impact
of gravity on the bubble results in a relative surface distortion
on the order of only $10\%$. We are interested in smaller bubbles
so that the distortion effects are insignificant.

Because of symmetry, we use the spherical coordinates $r$,
$\vartheta$, $\alpha$ with the origin in the center of the bubble.
As we have announced in Sec.~\ref{sec:intro}, we will demonstrate
that the shape oscillations are able to interact with the
breathing mode even within the linear approximation. Because in
this approximation the nonaxisymmetric modes of the shape
oscillations do not interact with the breathing mode (see
argumentation below in this section), we eventually restrict our
consideration to the axisymmetric problem.

We now formulate the dimensionless governing equations and
boundary conditions by measuring the distance, time, velocity
potential, deviations of pressure and the bubble surface in the
scales of $R$, $\sqrt{\rho R^3/\sigma}$, $\epsilon\sqrt{\sigma
R/\rho}$, $\epsilon \sigma/R$, and $\epsilon R$, respectively. As
one sees, the small parameter $\epsilon$ has the meaning of the
ratio of the amplitude of the surface oscillation to the
equilibrium radius $R$ of the bubble. Small oscillations of the
inviscid incompressible ambient are governed by the Bernoulli
equation and the condition of incompressibility:
\begin{equation} \label{sec:NO-goveq}
p=-\frac{\partial\varphi}{\partial t}, \quad \nabla^2 \varphi=0.
\end{equation}

The solid surface $\vartheta=\pi/2$ is impermeable for the liquid:
\begin{equation} \label{sec:NO-bc1}
\frac{\partial\varphi}{\partial \vartheta}=0.
\end{equation}
At the free surface, governed by the equation $r=1+\epsilon
\,\zeta(\vartheta,t)$, we prescribe the kinematic and the dynamic
conditions
\begin{eqnarray}
\frac{\partial \zeta}{\partial t}&=&\frac{\partial
\varphi}{\partial r}, \quad p-p_g = (\nabla^2_{\vartheta}+2)\zeta, \label{sec:NO-bc2} \\
\nabla^2_{\vartheta}&=&
\frac{1}{\sin\vartheta}\frac{\partial}{\partial \vartheta}\left(
\sin\vartheta \frac{\partial}{\partial \vartheta} \right),
\nonumber
\label{sec:NO-bc3}
\end{eqnarray}
\noindent where $\varphi$ is the velocity potential, $p$ and $p_g$
are the pulsation parts of the pressure in the liquid and gas
phases, respectively. These pulsation fields describe the pressure
deviation from their equilibrium values: $\epsilon^{-1}p_{g0}, \
p_{g0}\equiv P_g R/\sigma$ in the gas and
$\epsilon^{-1}\left(p_{g0}-2\right)$ in the liquid.

Next we define the spatially uniform oscillations of the gas
pressure $p_g$. Because the pulsations of the gas pressure in the
bubble are spatially homogeneous and the dissipative processes are
assumed to be negligible during a period of oscillation [recall
the accepted restrictions (\ref{sec:intro-nodamping})], we can
apply the adiabatic law for perfect gas. As a result, we arrive at
the condition
\begin{equation}
\left[p_{g0}+p_g(t)\right]V^\gamma(t)=p_{g0}V_0^\gamma. \nonumber
\end{equation}
\noindent Here $\gamma$ is the adiabatic exponent,
$V_0=\frac{2}{3}\pi$ is the dimensionless volume of motionless
bubble, and
\begin{equation}
V(t)=V_0\left(1+3\epsilon \left<\zeta\right>\right), \quad
\left<\zeta\right>=\frac{1}{2\pi}\int_S \zeta \,dS, \nonumber
\end{equation}
where the angle brackets denote the space averaging over
equilibrium surface $S=S(\vartheta)$ of the bubble. Thus, we
obtain for the pressure pulsation in the gas phase:
\begin{equation}\label{sec:NO-bc_pg}
p_g=-3\gamma p_{g0}\left<\zeta\right>\equiv
-\Pi_0\left<\zeta\right>,
\end{equation}
\noindent where we introduce parameter $\Pi_0=3\gamma \,P_g
R/\sigma$. Because $\Pi_0\propto p_{g0}$, hereafter $\Pi_0$ is
referred to as the dimensionless equilibrium pressure inside the
bubble. Note that this parameter can be presented as a ratio of
frequencies of the volume and shape oscillations squared,
$\Pi_0=\left(\omega_b \tau_c\right)^2$. Thus, one can clearly see
that for finite values of $\Pi_0$ inequalities
(\ref{sec:intro-times}) follow directly from requirements
(\ref{sec:intro-nodamping}).

We emphasize that for nonaxisymmetric modes
$\left<\zeta\right>=0$. As it follows from
Eq.~(\ref{sec:NO-bc_pg}), the pressure inside the bubble for these
modes does not change in time, $p_g(t)=0$, so that the breathing
mode cannot emerge. In other words, because in the linear problem
the modes with different azimuthal numbers are noninteracting, no
excitation of the volume oscillations is possible. Hence, to be
able to focus on the announced interaction between the breathing
mode and the shape oscillations hereafter we deal with the
axisymmetric problem, $\left<\zeta\right>\ne 0$.

The formulation of the boundary value problem is completed by
prescribing the dynamics of the contact line, where we impose the
Hocking condition (\ref{sec:intro-hocking})
\begin{equation} \label{sec:NO-bc4}
\frac{\partial\zeta}{\partial t}=-\lambda\frac{\partial
\zeta}{\partial \vartheta},
\end{equation}

\noindent where the dimensionless number
$\lambda=\Lambda\sqrt{\rho R/\sigma}$ is the Hocking (also known
as wetting) parameter.

The boundary value problem (\ref{sec:NO-goveq})-(\ref{sec:NO-bc4})
has been formulated in the linear approximation with respect to
small $\epsilon$. Particularly, this approximation allows us to
pose the boundary conditions (\ref{sec:NO-bc2}) at the
time-averaged, i.e., at the equilibrium, position of the surface,
$r=1$. The formulated governing equations and boundary conditions
involves the two dimensionless parameters $\Pi_0$ and $\lambda$.
The limiting case of high pressure, $\Pi_0 \to \infty$, refers to
the consideration of incompressible gas. In this situation, the
problem (\ref{sec:NO-goveq})-(\ref{sec:NO-bc4}) transforms to the
one describing the natural oscillations of an incompressible
bubble immersed in a liquid\cite{lyubimov-etal-04} (see a detailed
discussion in Sec.~\ref{sec:weak-compressibility}). Conversely,
for small values of $\Pi_0$ one expects that the bubble collapses.
Note that hereafter we assume the pressure in the ambient fluid to
be positive and are not interested in consideration of cavitation.
The limiting situations with respect to parameter $\lambda$
correspond either to the fixed contact line ($\lambda \to 0$, the
contact angle can change) or to the fixed contact angle ($\lambda
\to \infty$, the motion of the contact line is allowed). As it has
been mentioned before, apart from these particular situations the
Hocking condition leads to energy dissipation near the contact
line. For this reason, the natural oscillations are generally
damped.

Let us represent the decaying at the infinity solution to the
Laplace equation (\ref{sec:NO-goveq}), which satisfies the
impermeability condition (\ref{sec:NO-bc1}), in the form
\begin{equation} \label{sec:NO-potential}
\varphi={\rm Re}\left[ i\omega \, \sum_{n=0}^{\infty}\frac{A_n
P_{2n}\left(\theta\right)}{r^{2n+1}}e^{i \omega t}\right].
\end{equation}
\noindent Here we introduce the variable $\theta=\cos\vartheta$
and the Legendre polynomials $P_n(\theta)$.

Substitution of ansatz (\ref{sec:NO-potential}) into the Bernoulli
equation (\ref{sec:NO-goveq}) and kinematic condition [the first
relation in Eq.~(\ref{sec:NO-bc2})] leads to expressions
\begin{subequations}
\label{sec:NO-p-zeta}
\begin{eqnarray}
p & = & {\rm Re} \left[ \omega^2 \sum_{n=0}^{\infty}\frac{A_n P_{2n}(\theta)}{r^{2n+1}}e^{i \omega t}\right], \label{sec:NO-p} \\
\zeta & = & -{\rm Re} \left[ \sum_{n=0}^{\infty}(2n+1)\, A_n
P_{2n}(\theta)\,e^{i \omega t}\right], \label{sec:NO-zeta}
\end{eqnarray}
\end{subequations}
Next we determine $\zeta$ from the dynamic condition [the second
relation in Eq.~(\ref{sec:NO-bc2})]
\begin{equation}
\nonumber \zeta =  {\rm Re} \left[
\left\{\frac{1}{2}\Pi_0\left<\zeta \right>-\sum_{n=0}^{\infty}
\frac{\omega^2A_n P_{2n}(\theta)}{(2n-1)(2n+2)} + C\theta
\right\}\,e^{i \omega t}\right ],
\end{equation}
\noindent and equate it with Eq.~(\ref{sec:NO-zeta}). Accounting
for a relation $\left<\zeta\right>=-A_0$, we obtain the
coefficients introduced in (\ref{sec:NO-potential}):
\begin{equation} \label{sec:NO-An}
A_0=\frac{C}{\Omega_0^2-\omega^2}, \quad
A_n=\frac{(4n+1)P_{2n}(0)C}{\Omega_n^2-\omega^2} \;\; (n>0).
\end{equation}
\noindent Here, $\Omega_n^2=(2n-1)(2n+1)(2n+2)$ are the
eigenfrequencies of the shape oscillations of a spherical bubble.
These frequencies refer to the even modes; $\Omega_0^2=\Pi_0-2$ is
the frequency of the volume oscillations of a spherical bubble in
liquid. Here, the term $-2$ presents the well-known correction to
the frequency of breathing mode caused by surface
tension.\cite{plesset-prosperetti-77}

After the substitution of (\ref{sec:NO-p-zeta}) and
(\ref{sec:NO-An}) into condition (\ref{sec:NO-bc4}), we arrive at
the dispersion relation defining the spectrum of the
eigenfrequencies of a bubble:
\begin{eqnarray}
i\omega \left(\omega^2\; \sum_{n=1}^{\infty}\frac{\alpha_n
P_{2n}(0)}{\Omega_n^2 - \omega^2}-\frac{1}{2}-\frac{1}{\Omega_0^2-\omega^2}\right) & = & \lambda  \label{sec:NO-disp_relation} \\
\alpha_n=-\frac{(4n+1)P_{2n}(0)}{(2n-1)(2n+2)},&& \nonumber
\end{eqnarray}
\noindent where $\alpha_n$ are the coefficients in expansion of
$\theta$ in the series of the even Legendre polynomials at $\theta
\in [0,1]$.

Note that except for some particular situations (e.g., small and
high values of $\lambda$) the eigenfrequencies defined by
Eq.~(\ref{sec:NO-disp_relation}) are complex. Separating the real
and imaginary parts, one can demonstrate that the imaginary part
of the eigenfrequency is non-negative for any $\Pi_0>2$, i.e., the
oscillations decay. At $\Pi_0\approx 2$ and a finite value of the
wetting parameter, one of the eigenfrequencies is determined by
the relation
\begin{equation} \label{sec:NO-omega_pi2}
\omega=i\lambda(\Pi_0-2),
\end{equation}
which means that too small pressure inside the bubble results in
the monotonic instability of the bubble with respect to collapse.
Thus, relation (\ref{sec:NO-omega_pi2}) defines the stability
threshold against the fast (adiabatic) compression. Clearly, at
positive external pressure the bubble becomes unstable for higher
values of the gas pressure: $\Pi_0<6\gamma$ (or, in dimensional
units, $P_g<2\sigma/R$).

Consider now the dispersion relation in the limiting cases. For
the fixed contact angle, $\lambda \to \infty$, the eigenmodes of
oscillation of a hemispherical bubble coincide with the
corresponding even modes of the spherical bubble. In this case,
the oscillation frequencies equal $\Omega_k$ $(k=0, \, 1, \,
\dots)$. For high but finite values of the wetting parameter the
eigenfrequencies obey a relation
\begin{equation} \label{sec:NO-omega}
\omega^{(k)}=\Omega_k+\frac{i\gamma_k}{\lambda}+\frac{\gamma_k\Omega_k}{2\lambda^2}\left(\sum_{
\begin{array}{cc}
n=0, \\[-1mm] n\neq k
\end{array}
}^{\infty}\frac{4\gamma_n}{\Omega_n^2-\Omega_k^2}-\frac{\gamma_k}{\Omega_k^2}\right),
\end{equation}
\noindent defined for $k=0, \, 1, \, \dots$. Here,
$\gamma_k=-\Omega_k^2\alpha_k P_{2k}(0)/2$ $(k>0)$,
$\gamma_0=1/2$. As it can be seen, the oscillations are weakly
damped, a correction to the eigenfrequency is proportional to
$\lambda^{-2}$. Formula (\ref{sec:NO-omega}) is identical to that
for the incompressible bubble for positive $k$, if the first term
in the sum is vanishing, which corresponds to the limit
$\Omega_0^2\to\infty$. We indicate that result
(\ref{sec:NO-omega}) holds even for $\Pi_0<2$. In this case all
the terms for $k=0$ are imaginary and relation
(\ref{sec:NO-omega}) describes two branches in the spectrum (the
stable and unstable): $\omega^{(0)}=\pm i
\sqrt{2-\Pi_0}+O(\lambda^{-1})$.

The frequency of oscillation of the bubble with the fixed contact
line $\omega_p$ is determined from the following real expression:
\begin{equation} \label{sec:NO-omega_p}
\omega_p^2 \sum_{n=1}^{\infty} \frac{\alpha_n
P_{2n}(0)}{\Omega_n^2-\omega_p^2}-\frac{1}{2}-\frac{1}{\Omega_0^2-\omega_p^2}=0.
\end{equation}

In the case of $\lambda \ll 1$ (slight slip of the contact line)
the results are qualitatively similar to those of incompressible
liquid:\cite{lyubimov-etal-06, lyubimov-etal-04} the decay rate is
proportional to the small wetting parameter and a correction to
the frequency $\omega_p$ is proportional to $\lambda^2$.

However, despite the qualitative similarity with the oscillations
of the hemispherical drop studied in
Refs.~\onlinecite{lyubimov-etal-06, lyubimov-etal-04}, there
arises a number of peculiar effects for the compressible bubble.
The appearance of an additional branch in the spectrum of natural
oscillation, governed by the parameter $\Pi_0$, leads to a
nontrivial effect: the volume and the shape oscillations start to
interact. Thus, if for a spherical bubble of the same radius the
frequency $\Omega_0$ of the volume oscillations is close to any of
the frequencies $\Omega_k$ of the shape oscillations (for an even
mode)
\begin{equation}
\Omega_0^2=\Omega_k^2(1+\delta), \quad \delta \ll 1, \nonumber
\end{equation}
\noindent then one of the eigenfrequencies $\omega_{0k}$ for the
hemispherical bubble is in between these frequencies:
\begin{equation} \label{sec:NO-omega_0k}
\omega_{0k}=\frac{\gamma_0\Omega_k}{\gamma_0+\gamma_k}+\frac{\gamma_k\Omega_0}{\gamma_0+\gamma_k}.
\end{equation}
%
%
\begin{figure}[!t]
\centering
\includegraphics[width=0.44\textwidth]{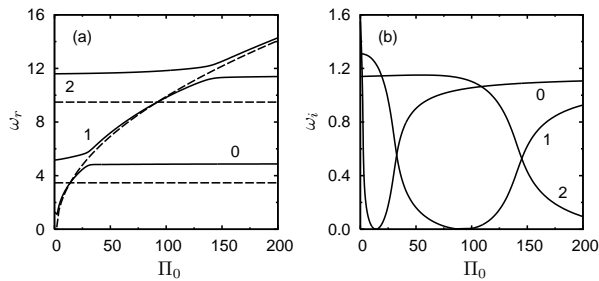}\vspace{-2mm}
\caption{Eigenfrequencies $\omega$ of the bubble oscillations as
functions of pressure $\Pi_0$ for modes $0-2$. The
real (a) and imaginary (b) parts of $\omega$ are plotted for
$\lambda=1$ (full line) and $\lambda=10$ (dashed line).}
\label{fig2}
\end{figure}
%

A complex correction to the frequency defined by
(\ref{sec:NO-omega_0k}) is proportional to $\delta^2$. For this
reason, the decay is weak for this mode irrespective of the
wetting parameter and in the ``resonant'' situation ($\delta=0$)
is absent at all. With the accuracy up to the terms of order
$\delta^2$ only two coefficients, $A_0$ and $A_k$, are
nonvanishing in sums (\ref{sec:NO-potential}) and
(\ref{sec:NO-p-zeta}). Hence, this mode is a superposition of two
kinds of motion: the radial pulsations and the $k$-th mode of the
shape oscillations of the bubble with the fixed contact angle.
These oscillations occur in the antiphase and their relative
amplitudes are such that the contact line remains motionless.

For arbitrary values of the governing parameters,
Eq.~(\ref{sec:NO-omega}) was solved numerically. Using the secant
method we retained up to 200 terms. Careful analysis has shown
that for most situations retention of only 20 terms is enough to
ensure convergence.

In Fig.~\ref{fig2} we present the frequency of natural
oscillations as a function of the pressure in the bubble. At
$\lambda=10$ the real part of the frequencies of the first modes
almost coincide with the values $\Omega_0$, $\Omega_1$,
$\Omega_2$. The imaginary part of the frequency is small and is
described by formula (\ref{sec:NO-omega}). A distinction exists
only in the case $\Omega_0\approx\Omega_k$: the imaginary part of
the frequency drastically decreases for the volume oscillations
and increases for the shape oscillations. At moderate values of
the wetting parameter ($\lambda=1$), with the growth of $\Pi_0$
the spectrum of the eigenfrequencies rearranges. In this case the
damping time of oscillations is comparable with their period; the
imaginary part of the frequency as a function of $\Pi_0$ is given
in Fig.~\ref{fig2}(b). As has been indicated before, near the
``resonance'' $\Omega_0\approx\Omega_k$ the imaginary part of the
frequency for one of the modes tends to zero according to the law
$\omega_i \simeq \delta^2 \propto (\Omega_0-\Omega_k)^2$.

%
%
\begin{figure}[!t]
\centering
\includegraphics[width=0.44\textwidth]{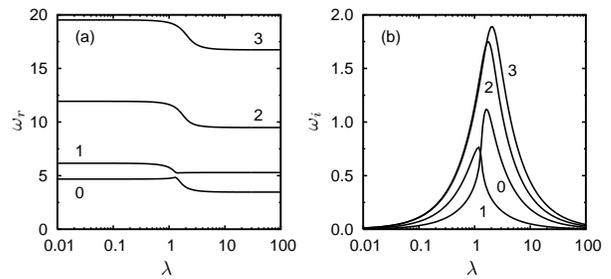}\vspace{-2mm}
\caption{Eigenfrequencies $\omega$ of the bubble oscillations as
functions of wetting parameter $\lambda$ for modes $0-3$. The real (a) and imaginary (b) parts of $\omega$ are
plotted for $\Pi_0=30$.} \label{fig3}
\end{figure} %
%
%
%
\begin{figure}[!b]
\centering
\includegraphics[width=0.22\textwidth]{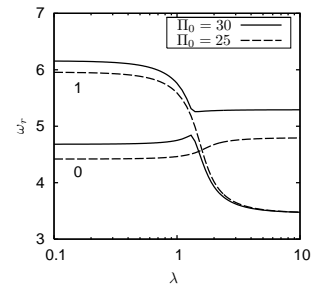}\vspace{-2mm} 
\caption{Rearrangement of the two lowest modes  $0$ and $1$ under
small variation of pressure $\Pi_0$. Eigenfrequencies $\omega_r$
as functions of wetting parameter $\lambda$, plotted for
$\Pi_0=30$ and $\Pi_0=25$, cf. Fig.~\ref{fig3}(a). \vspace{-5mm}} \label{fig4}
\end{figure} %
%

The dependence of eigenfrequencies on the Hocking parameter at a
fixed gas pressure in the bubble can be found in Fig.~\ref{fig3}.
For modes $2$ and $3$, the behavior resembles the case of the
incompressible liquid:\cite{lyubimov-etal-06,lyubimov-etal-04} the
real part decreases with the growth of $\lambda$, the imaginary
part has a maximum at a finite value of the wetting parameter,
turning to zero at small and high $\lambda$. An interesting
behavior of the eigenfrequencies can be observed for the two
lowest modes, when the complex frequencies become close, see
Fig.~\ref{fig4}. Note that an insignificant change of $\Pi_0$ is
able to qualitatively rearrange the dependence
$\omega_r(\lambda)$. We indicate that the exact coincidence of
complex decay rates of the two modes resulting in such a
rearrangement of the spectrum is possible only in a discrete
number of points $(\lambda, \Pi_0)$, which obey the complex-valued
equation $\omega^{(k)}(\lambda,\Pi_0)=\omega^{(n)}(\lambda,\Pi_0)$. \\

As the final point of this Section, we point out to what
extent the presence of the substrate influences damping of a
bubble oscillation. Generally (see, e.g., Refs.~ \onlinecite{wijngaarden-72,
nigmatulin-91}), oscillation damping of a spherical bubble may be
caused by viscosity, thermal diffusion, and radiation of acoustic
wave. Although all these phenomena are relevant for the
hemispherical bubble, they can be remarkably changed by the
substrate.

As we stressed in Sec.~\ref{sec:intro}, understanding the impact
of viscosity on the moving contact line
is a notoriously complicated problem. However, an important
estimation can be drawn from the problem with the contact line
pinned,\cite{lyubimov-etal-06} $\lambda=0$. According to this
analysis, the dimensionless decay rate is
$\sqrt{\tau_c/\tau_v}$ against conventional $\tau_c/\tau_v$
for a spherical bubble away from the solid surface.

A similar situation is expected for the damping caused by heat
diffusion. Note that this mechanism may be important for both gas
and fluid. Moreover, the ratios of thermal conductivities of the three media
(gas, liquid, and solid) become important. If the conductivity of
the solid is not negligibly small, the thermal boundary layer near
the solid surface is developed. The decay rate is given by
$\sqrt{\omega_b\tau_t}$, whereas for a bubble in the absence of
the solid surface one ends up with the usual value of
$\omega_b\tau_t$.

Acoustic irradiation has nondissipative origin and therefore the
presence of the substrate is insignificant. Hence, the
conventional condition
$\omega_b \tau_a \ll 1$ as in (\ref{sec:intro-nodamping}) is
sufficient to neglect this phenomenon.

As we see, the presence of the solid surface leads to the
development of the boundary layers, which results in the faster
oscillation damping compared to the case of a bubble away from the
substrate. These changes, however, do not change restrictions
(\ref{sec:intro-nodamping}). Thus, the criteria that allow us to
neglect viscous and heat dissipation and acoustic irradiation loss
remain conventional.

\section{Forced oscillations} \label{sec:forced-oscillations}

%
%
\begin{figure}[!t]
\centering
\includegraphics[width=0.28\textwidth]{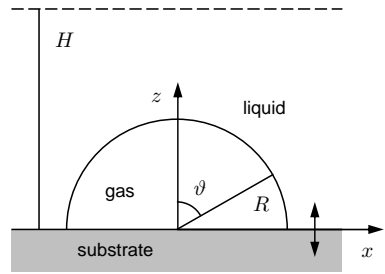}
\caption{The geometry of the forced oscillations problem.} \label{fig5}
\end{figure} %
%
%
Consider the behavior of a gas bubble sitting on a plane solid
substrate, as sketched in Fig.~\ref{fig5}. We neglect gravity as
before and now look at the problem of forced oscillations. Assume
that the substrate performs transversal vibrations (with respect
to its plane) with the amplitude $a$ and the frequency $\omega$;
the substrate velocity in an inertial reference frame is
$a\omega\sin\omega t$. We consider the frequency of vibrations to
be comparable with both the frequencies of the shape and volume
oscillations. We stress that spatially uniform pulsations are of
crucial importance for the consideration of the bubble. This is in
contrast to the problems addressing an incompressible
drop,\cite{lyubimov-etal-06, lyubimov-etal-04} where the system is
dominantly governed by the pressure difference in the liquid. In
our system, to define the uniform part of the pressure field one
has to specify an additional condition away from the bubble. From
the experimental point of view, the most convenient way to
overcome this difficulty is to attach the bubble to the bottom of
a liquid layer of depth $H$ (large compared to the size of bubble)
with the free surface,\cite{thanks-DV} see Fig.~\ref{fig5}. For a
spherical bubble, a similar way of treating this peculiarity of
the pressure field has been applied before,\cite{nigmatulin-91,
grigoryan-etal-65} where the bubble has been immersed in a
vibrated column of liquid with a free surface and the impact of
vibrations on the spherically symmetric oscillation mode has been
studied.

In the reference frame moving together with the substrate the
small oscillations of the bubble are governed by the following
dimensionless equations and boundary conditions:
\begin{subequations}
\label{sec:FO-eqs-bcs}
\begin{eqnarray}
p & = & -\frac{1}{\Omega^2}\frac{\partial \varphi}{\partial
t}+\left(1-\frac{z}{h}\right)\cos\Omega t, \quad \nabla^2 \varphi=0, \label{sec:FO-goveq} \\
\vartheta & = & \frac{\pi}{2}: \; \frac{\partial
\varphi}{\partial \vartheta} = 0, \label{sec:FO-bc1} \\
r & = & 1: \; \frac{\partial \zeta}{\partial t} =
\frac{\partial \varphi}{\partial r}, \; \Omega^2p+\Pi_0\left<\zeta\right>=\left(\nabla^2_{\vartheta}+2\right)\zeta, \label{sec:FO-bc2} \\
z & = & h: \; p=0, \label{sec:FO-bc3} \\
r & = & 1, \; \vartheta=\frac{\pi}{2}: \; \frac{\partial
\zeta}{\partial t} = - \lambda \frac{\partial \zeta}{\partial
\vartheta}. \label{sec:FO-bc4}
\end{eqnarray}
\end{subequations}

The problem (\ref{sec:FO-eqs-bcs}) has been nondimensionalized
using the scales $aH/R$, $\rho a \omega^2 H$, $aH\sqrt{\sigma/\rho
R^3}$ for the deviation of the bubble surface from its equilibrium
form, the pressure, and the velocity potential, respectively; for
the distance and time we use the same scales as before, in problem
(\ref{sec:NO-goveq})-(\ref{sec:NO-bc4}). We note that the oscillations of the
bubble surface can be considered to be small only provided that
\begin{equation} \label{sec:FO-ampl-restr}
a \ll \frac{R^2}{H}.
\end{equation}
\noindent As we see, the restriction imposed on the amplitude $a$
is much stricter than in the situation of an incompressible drop
($a \ll R$). This restriction, however, can be weakened for the
weakly compressible bubble, see
Sec.~\ref{sec:weak-compressibility}.

Among the dimensionless parameters $\Pi_0$ and $\lambda$, defined
earlier, there appear two more parameters in problem
(\ref{sec:FO-eqs-bcs}): the dimensionless frequency $\Omega$ of
the substrate oscillation, which is related to the Weber number
$\Omega^2=\rho \omega^2 R^3/\sigma$, and the relative width of the
layer $h=H/R$. As we have assumed above, the last parameter is
high:
\begin{equation} \label{sec:FO-high-h}
h \gg 1.
\end{equation}
\noindent This inequality allows us to neglect the term
proportional to $z/h$ in the Bernoulli equation
(\ref{sec:FO-goveq}), and apply an approximation. We replace the
exact condition (\ref{sec:FO-bc3}) for the pressure at the free
surface with a requirement
\begin{equation} \label{sec:FO-phi-zero}
r \to \infty, \quad \varphi\to 0,
\end{equation}
\noindent which ensures the disturbance decay far from the bubble.

Indeed, in the absence of the bubble the vibrations of the layer
cause pulsations of the pressure in the liquid:
$p_0=(1-z/h)\cos\omega t$. We note that near the bottom, where $z
\ll h$, the inertial force is negligible, therefore these
pulsations can be considered as spatially uniform. Thus, far away
from the bubble but close to the solid surface, the time
oscillations of the pressure are spatially uniform, which causes
the volume (and hence also the shape) oscillations of the bubble.
Surely, the wave scattered by the bubble does not satisfy
condition (\ref{sec:FO-bc3}) precisely. This solution, however,
can be easily corrected by the method of images. Such a procedure
introduces an error of order $h^{-1}$ in the condition for the
pressure at the bubble surface (\ref{sec:FO-bc2}), which is of the
same order compared with the neglected inertial term.

The solution to the Laplace equation for the velocity potential
that decays at infinity and satisfies the impermeability condition
(\ref{sec:FO-bc1}) along with the fields of pressure and surface
deviation, can be represented as follows:
\begin{subequations}
\label{sec:FO-ansatzs}
\begin{eqnarray}
\varphi & = & {\rm Re} \left[ i\Omega \sum_{n=0}^{\infty}\frac{A_n
P_{2n}(\theta)}{r^{2n+1}}e^{i\Omega t} \right],
\label{sec:FO-ansatz-varphi} \\
p & = & {\rm Re} \left[ \left(\sum_{n=0}^{\infty}\frac{A_n
P_{2n}(\theta)}{r^{2n+1}}+1 \right)e^{i\Omega t} \right],
\label{sec:FO-ansatz-p} \\
\zeta & = & -{\rm Re} \left[\sum_{n=0}^{\infty} (2n+1) A_n
P_{2n}(\theta)e^{i\Omega t}\right]. \label{sec:FO-ansatz-zeta}
\end{eqnarray}
\end{subequations}
\noindent Applying the dynamic boundary condition, we obtain the
expansion coefficients
\begin{equation} \label{sec:FO-An}
A_0=\frac{\Omega^2+C}{\Omega_0^2-\Omega^2}, \quad
A_n=\frac{(4n+1)P_{2n}(0)C}{\Omega_n^2-\Omega^2} \;\;\; (n>0).
\end{equation}
Here, the complex constant $C$ is found from the Hocking condition
(\ref{sec:FO-bc4}):
\begin{equation} \label{sec:FO-C}
C=\Omega^2\left[\left(\Omega_0^2-\Omega^2\right)\left(
\Omega^2\sum_{n=1}^{\infty}\frac{\alpha_n
P_{2n}(0)}{\Omega_n^2-\Omega^2} - \frac{1}{2} +
\frac{i\lambda}{\Omega} \right) -1 \right]^{-1}.
\end{equation}
As it can be seen, $\arg(A_n)$ is nonzero and identical for any
$n>0$, but different from $\arg(A_0)$. This indicates that the
bubble oscillations is a superposition of the standing wave (shape
oscillations) and the volume oscillations, which have phase shifts
relative to each other and relative to the substrate vibrations.

Further, one can show\cite{lyubimov-etal-06, lyubimov-etal-04}
that the frequencies $\Omega_k$ are not resonant at any value of
$\lambda$. At $\Omega=\Omega_k$ the motion of the bubble
represents a combination of the radial (volume) oscillation and
the $k$-th mode of the shape oscillations (both these oscillations
are in phase with the vibration motion of the substrate):
\begin{equation}\label{sec:FO-zeta}
\zeta=\frac{\Omega_k^2}{\Omega_0^2-\Omega_k^2}\left(
\frac{P_{2k}(\theta)}{P_{2k}(0)}-1 \right)\cos\Omega_k t \;\;\;
(k>0),
\end{equation}
\noindent which means that the contact line remains motionless.

At a frequency close to $\Omega_0$, we put
$\Omega=\Omega_0(1+\delta_0)$, where the frequency mismatch
$\delta_0 \ll 1$, and obtain from (\ref{sec:FO-An}),
(\ref{sec:FO-C}):
\begin{eqnarray}
A_0 &\approx& -\Omega_0^2 \left(
\Omega_0^2\sum_{n=1}^{\infty}\frac{\alpha_n
P_{2n}(0)}{\Omega_n^2-\Omega_0^2} - \frac{1}{2} +
\frac{i\lambda}{\Omega_0} \right), \nonumber \\
A_n&\approx&-\frac{(4n+1)P_{2n}(0)\Omega_0^2}{\Omega_n^2-\Omega_0^2}
\;\;\; (n>0), \nonumber \\
C&\approx& -\Omega^2\left[1-\delta_0\Omega_0^2\left(
\Omega_0^2\sum_{n=1}^{\infty}\frac{\alpha_n
P_{2n}(0)}{\Omega_n^2-\Omega_0^2} - \frac{1}{2} +
\frac{i\lambda}{\Omega_0} \right) \right]. \nonumber
\end{eqnarray}
\noindent Performing evaluation for $\zeta_0=\zeta(\theta=0)$ and
accounting for a relation $\sum_{n=0}^{\infty}\alpha_n
P_{2n}(0)=0$, we end up with
%
%
\begin{equation}
\zeta_0\approx {\rm Re} \left[i \lambda \Omega_0 e^{i\Omega_0
t}\right] =-\lambda \Omega_0 \sin \Omega_0 t. \label{sec:FO-zeta0}
\end{equation}
\noindent This result clearly indicates that at a finite value of
$\lambda$ and a vibration frequency close to the frequency of the
breathing mode, the bubble oscillates with a finite amplitude.

Next, in the limiting case of the fixed contact angle ($\lambda
\gg 1$) we arrive at an obvious conclusion: at any frequency the
bubble performs radial oscillations with an amplitude
\begin{equation}
A_0=\frac{\Omega^2}{\Omega^2_0-\Omega^2}. \label{sec:FO-A0}
\end{equation}
\noindent Other coefficients in (\ref{sec:FO-ansatzs}) are small
(of order $\lambda^{-1}$) because the contact line only weakly
interacts with the substrate. If, however, the frequency $\Omega$
of the external force is close to the frequency $\Omega_k$ of the
$k$-th mode of the shape oscillations, then the resonant
amplification of this mode occurs:
\begin{eqnarray}
\zeta & = & -b_0\left[\cos\Omega_k t - a_k
\frac{P_{2k}(\theta)}{P_{2k}(0)}\cos(\Omega_k t + \beta_k)
\right], \label{sec:FO-zeta-shape-res} \\
a_k & = &
\frac{\gamma_k}{\sqrt{(\Omega_k-\Omega)^2\lambda^2+\gamma_k^2}},
\; \tan\beta_k=\frac{(\Omega_k-\Omega)\lambda}{\gamma_k},
\label{sec:FO-ak-res} \nonumber
\end{eqnarray}
\noindent where $b_0=\Omega_k^2/(\Omega_0^2-\Omega_k^2)$. Thus,
close to the resonant frequency (for the fixed contact angle) the
bubble motion consists of a superposition of the radial
oscillation and a standing wave (one mode in the expansion), which
have a relative phase shift. Exactly at the point of resonance the
solution is given by formula (\ref{sec:FO-zeta}).

For large $\lambda$, in the situation where the frequency of
oscillations is close to the eigenfrequency of the volume
oscillations, $\Omega\approx \Omega_0$, a resonant amplification
of the radial pulsations takes place:
\begin{eqnarray}
\zeta & = & a_0\cos(\Omega_0 t+\beta_0), \label{sec:FO-zeta-vol-res} \\
a_0 & = & \frac{\lambda \gamma_0
\Omega_0}{\sqrt{(\Omega_0-\Omega)^2\lambda^2+\gamma_0^2}}, \;
\tan\beta_0=\frac{\gamma_0}{(\Omega-\Omega_0)\lambda}. \nonumber
\label{sec:FO-a0-res}
\end{eqnarray}
\noindent Note that exactly at the point of resonance,
$\Omega=\Omega_0$, the obtained result (\ref{sec:FO-zeta-vol-res})
is reduced to Eq.~(\ref{sec:FO-zeta0}), which means that the two
asymptotics match in the overlapping range of parameters:
$\Omega\approx \Omega_0$ and $\lambda \gg 1$. On the other hand,
the amplitudes in (\ref{sec:FO-zeta-vol-res}) and relation
(\ref{sec:FO-A0}) become the same for $\Omega_0 \gg
|\Omega-\Omega_0| \gg \lambda^{-1}$.

We also indicate that resonant solution
(\ref{sec:FO-zeta-vol-res}) is characterized by oscillations with
the amplitude $a_0=O(\lambda)$, which is much higher than the
resonant amplitude $a_k$ at a frequency of the shape oscillations.
As it follows from the resonant solution
(\ref{sec:FO-zeta-shape-res}), the amplitude $a_k \simeq O(1)$ and
hence $a_0/a_k \sim \lambda \gg 1$.
%
%
\begin{figure}[!tb]
\centering
\includegraphics[width=0.44\textwidth]{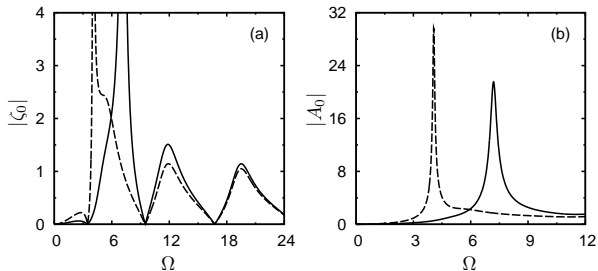}\vspace{-2mm}
\caption{Amplitude-frequency response for the contact line
$\zeta_1$ (a) and volume (b) oscillations evaluated for
$\lambda=1$ and pressure values $\Pi_0=50$ (full line) and
$\Pi_0=20$ (dashed line).} \label{fig6}
\end{figure} %
%

We now consider the bubble dynamics when the three frequencies are
close: $\Omega\approx\Omega_0\approx\Omega_k$. For arbitrary
values of $\lambda$ we obtain
\begin{equation}
\zeta=A\left( 1-\frac{P_{2k}(\theta)}{P_{2k}(0)}
\right)\cos\omega_k t, \label{zeta_double}
\end{equation}
\noindent i.e., the bubble oscillations are in phase with the
substrate vibrations and the contact line is motionless. The
amplitude of oscillations read
\begin{equation}
A=\frac{\gamma_0\gamma_k}{\gamma_0+\gamma_k}\,
\frac{\Omega_k}{\Omega-\omega_{0k}}, \label{sec:FO-A}
\end{equation}
\noindent where $\omega_{0k}$ is the eigenfrequency of
oscillations for the regime $\Omega_0\approx\Omega_k$, described
by Eq.~(\ref{sec:NO-omega_0k}). A more rigorous analysis indicates
that the phase of the resonant oscillations is shifted by $\pi/2$
with respect to the substrate motion, and their amplitude is
\begin{equation}
A_{res}=\frac{\Omega_k}{\lambda}\left(\frac{\gamma_0+\gamma_k}{\gamma_0}\right)^2\delta^{-2},
\quad \delta=\frac{\Omega^2_0-\Omega^2_k}{\Omega_k^2}.
\label{sec:FO-A-res}
\end{equation}
The resonant amplitude remains bounded even at the eigenfrequency
$\omega_{0k}$, except for the case of precise coincidence of all
the three frequencies. Note, the fact that the amplitude of the
bubble oscillations is inversely proportional to the frequency
mismatch squared $\delta^2$ is in agreement with the result
obtained in Sec.~\ref{sec:natural-oscillations}: at
$\Omega_0\approx\Omega_k$ energy dissipation is proportional to
$\delta^2$.

At arbitrary values of the governing parameters series
(\ref{sec:FO-ansatzs}) were evaluated numerically. In
Fig.~\ref{fig6} we present the amplitudes of the contact line
oscillations $\zeta_0\equiv\zeta(\theta=0)$ and the radial
pulsation as functions of frequency $\Omega$ for $\lambda=1$.
Because of the dissipative processes at the contact line, the
amplitude of resonant oscillations remains bounded. However, the
interaction of the volume and the shape oscillations leads to
considerable increase of the amplitude near the frequency defined
by relation (\ref{sec:NO-omega_0k}).

As it follows from solution (\ref{sec:FO-zeta}), the contact line
is fixed at the frequencies $\Omega$ coinciding with $\Omega_k$,
the eigenfrequencies of the bubble oscillations with the fixed
contact angle. At $\Omega=\omega_p$ (recall that $\omega_p$ is the
eigenfrequency of a bubble with the fixed contact line) the
amplitude of the contact line motion does not depend on the
Hocking parameter:
\begin{equation}\label{z0_om_p}
\zeta_{0}(\Omega=\omega_p)=\frac{\omega_p^2}{\omega_p^2-\Omega_0^2}.
\end{equation}
Note that for sufficiently small $\lambda$, the function
$\zeta_0(\Omega)$ possesses a local maximum at $\Omega=\omega_p$.
Moreover, the local maximum is rather close to $\Omega=\omega_p$
even at $\lambda=1$, see Fig.~\ref{fig6}. For instance, the second
maximum for $\Pi_0=20$ (the dashed line) takes place at
$\Omega_{max}=11.89$, whereas $\omega_p=11.94$.

%
%
\begin{figure}[!t]
\centering
\includegraphics[width=0.44\textwidth]{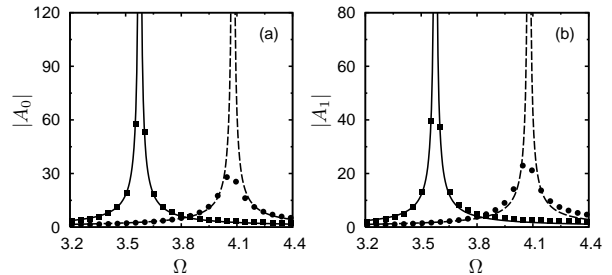}\vspace{-2mm}
\caption{Absolute values of coefficients $A_0$ (a) and $A_1$ (b)
as functions of $\Omega$ near the double resonance, $\lambda=1$.
Full and dashed lines correspond to Eqs.~(\ref{zeta_double}) and
(\ref{sec:FO-A}) at $\Pi_0=15$ ($\delta$=1/12) and $\Pi_0=20$
($\delta$=1/2), respectively. Squares ($\Pi_0=15$) and circles
($\Pi_0=20$) present numerically obtained results according to
Eqs. (\ref{sec:FO-An}) and (\ref{sec:FO-C}).} \label{fig7}
\end{figure} %
%

As it becomes clear from Fig.~\ref{fig7}, formulas
(\ref{zeta_double}) and (\ref{sec:FO-A}) work remarkably well at
significant deviations from the ``double resonance,'' e.g., the
amplitude of the radial pulsations at $\Pi_0=20$ ($\delta=-1/2$)
is still in good agreement with (\ref{sec:FO-A}).

\section{Oscillations of a weakly compressible bubble} \label{sec:weak-compressibility}

We now turn to the consideration of the weakly compressible
bubble, which implies high gas pressure, $\Pi_0 \approx \Omega_0^2
\gg 1$. Compared to the situation analyzed in
Sec.~\ref{sec:forced-oscillations}, the high pressure in the
bubble means that the radial pulsations as well as the induced
shape oscillations become small. Mathematically, these facts are
reflected by the smallness of $\zeta$ in
(\ref{sec:FO-ansatz-zeta}): as it can be seen from
Eq.~(\ref{sec:FO-An}), all the coefficients $A_n\to 0$ as
$\Omega_0 \to \infty$. What is important, the condition $\Pi_0 \gg
1$ does not guarantee against the insignificance of the bubble
compressibility. Indeed, the amplitude of the uniform part $p_0$
of pressure oscillations is based on the width of the layer: $\rho
a \omega^2 H$. This pressure contribution causes surface
deviations proportional to $\Pi_0^{-1}$, it is of crucial
importance for the bubble as a compressible object. Another source
of bubble oscillations comes from the inertial force and induces
bubble distortion independent of $\Pi_0$. The corresponding
contribution in the pressure, $p_{in}$, is important for any
bubble, does not matter compressible or not. This nonuniform part
of pressure $p_{in}\sim \rho a \omega^2 R$, produced by the
inertial force, is much smaller than the uniform part, because
$p_{in}/p_0\simeq h^{-1} \ll 1$. Thus, the dynamics of the weakly
compressible bubble is determined by the competition of the two
different factors: the weak compressibility itself ($\Pi_0^{-1}
\ll 1$) and the smallness of the inertial part of pressure
($h^{-1} \ll 1$).

Formally, there appear two small parameters in the problem:
$h^{-1}$ and $\Pi_0^{-1}$. Their ratio, the parameter that
describes the impact of compressibility on the bubble
oscillations, is assumed to be finite. Note that restriction
(\ref{sec:FO-ampl-restr}) imposed on the amplitude in
Sec.~\ref{sec:forced-oscillations} becomes much milder (see also
Ref.~\onlinecite{vibro-06}):
\begin{equation}
a \ll R, \quad a \ll \frac{\Pi_0}{h}R.
\end{equation}
Thus, the solution to problem (\ref{sec:FO-eqs-bcs}) can be
presented as
\begin{subequations}
\label{sec:WC-ansatz12}
\begin{eqnarray}
\varphi & = & \Pi_0^{-1}\varphi_1+h^{-1}\varphi_2,  \\
p & = & \cos\Omega t+\Pi_0^{-1}p_1+h^{-1}p_2,  \\
\label{zeta_weakly}\zeta & = & \Pi_0^{-1}\zeta_1+h^{-1}\zeta_2.
\end{eqnarray}
\end{subequations}
\noindent Here, the fields $\varphi_1$, $p_1$, $\zeta_1$ describe
the motion caused by the spatially uniform part of the pressure
and $\varphi_2$, $p_2$, $\zeta_2$ present the contribution induced
by the inertial force. The latter contribution is related to the
dynamics of an incompressible bubble. We also emphasize that
the accepted in Sec.~\ref{sec:natural-oscillations} condition
${\rm Bo} \ll 1$ is sufficient to neglect gravity here, for the
case of weakly compressible bubble. The smallness of the Bond
number ensures that the capillary effects produced by the inertial
force dominate over the contributions caused by gravity.

The solution to the first problem, for the fields $\varphi_1$,
$p_1$, $\zeta_1$, is obtained from the results of
Sec.~\ref{sec:forced-oscillations} in the limit $\Pi_0 \gg 1$
($\Omega_0 \gg 1$). Writing down the fields of the velocity
potential $\varphi_1$, pressure $p_1$, and surface deviation
$\zeta_1$ is terms of series (\ref{sec:FO-ansatzs}), we obtain
from (\ref{sec:FO-An}) the coefficients
\begin{subequations}
\label{sec:WC-cffs1}
\begin{eqnarray}
A_0 & = & \Omega^2, \;\;
A_n=\frac{(4n+1)P_{2n}(0)C_1}{\Omega_n^2-\Omega^2} \;\; (n>0), \\
C_1 & = & \Omega^2\left[ \Omega^2\sum_{n=1}^{\infty}
\frac{\alpha_n P_{2n}(0)}{\Omega_n^2-\Omega^2}
-\frac{1}{2}+\frac{i\lambda}{\Omega} \right]^{-1}.
\end{eqnarray}
\end{subequations}

\begin{figure}[!b]
\centering
\includegraphics[width=0.48\textwidth]{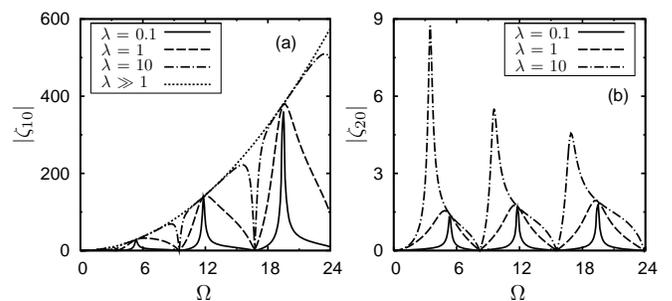}
\caption{Amplitude-frequency response for the contact line of the
weakly compressible bubble, $\zeta_1$ (a) and $\zeta_2$ (b),
evaluated for different values of $\lambda$. The asymptotics of
large $\lambda$ (left panel) is given by formula
$\zeta_1=-\Omega^2$.} \label{fig8}
\end{figure}

In Fig.~\ref{fig8}(a) we present the amplitude of the contact line
oscillations $\zeta_{10}\equiv \zeta_1(\theta=0)$ as a function of
frequency $\Omega$ for different $\lambda$. It is clearly seen
that at a fixed value of $\Omega$ this amplitude grows with the
increase of $\lambda$. This solution is closely related to the
mode discussed in Sec.~\ref{sec:forced-oscillations}.
Particularly, $\zeta_{10}=0$ at $\Omega=\Omega_k$ [see
Eq.~(\ref{sec:FO-zeta})]. At $\Omega=\omega_p^{(\infty)}$, where
$\omega_p^{(\infty)}$ is $\omega_p$ evaluated at the limit
$\Pi_0\to \infty$, the real amplitude of the contact line oscillations
is independent of $\lambda$ and equals $\Omega^2$ [cf.
Eq.~(\ref{z0_om_p}) at $\Omega_0^2\approx\Pi_0\gg 1$, recall the
weight factor $\Pi_0^{-1}$ in Eq.~(\ref{zeta_weakly})]. At
$\lambda \gg 1$ we conclude from Eqs.~(\ref{sec:WC-cffs1}) that
$C_1$ is small and $\zeta_1\approx -A_0=-\Omega^2$. Note that this
asymptotics works well even at $\lambda=10$ [see
Fig.~\ref{fig8}(a)], except for the close vicinity of
``antiresonant'' frequencies $\Omega=\Omega_k$, at which the
contact line is motionless.

%
%
%

The second problem, for $\varphi_2$, $p_2$, $\zeta_2$, is governed
by the following equations and boundary conditions:
\begin{subequations}
\label{sec:WC-problem2}
\begin{eqnarray}
p_2 & = & -\frac{1}{\Omega^2}\frac{\partial \varphi_2}{\partial
t}- z \cos\Omega t, \quad \nabla^2 \varphi_2=0, \label{sec:WC-goveq} \\
\vartheta & = & \frac{\pi}{2}: \; \frac{\partial
\varphi_2}{\partial \vartheta} = 0, \label{sec:WC-bc1} \\
r & = & 1: \; \frac{\partial \zeta_2}{\partial t}= \frac{\partial
\varphi_2}{\partial r}, \quad
\Omega^2 p_2 = \left(\nabla^2_{\vartheta}+2\right)\zeta_2, \label{sec:WC-bc2} \\
r & \to & \infty: \; \varphi_2 = 0, \label{sec:WC-bc3} \\
r & = & 1, \; \vartheta=\frac{\pi}{2}: \; \frac{\partial
\zeta_2}{\partial t} =  - \lambda \frac{\partial \zeta_2}{\partial
\vartheta}. \label{sec:WC-bc4}
\end{eqnarray}
\end{subequations}

Problem (\ref{sec:WC-problem2}) describes the forced oscillations
of an incompressible bubble immersed in a liquid. Qualitatively,
this problem is similar to that of the forced oscillations of a
hemispherical drop.\cite{lyubimov-etal-06} Hence, the solution can
be represented as
\begin{subequations}
\label{sec:WC-ansatz2}
\begin{eqnarray}
\varphi_2 & = & {\rm Re} \left[ i\Omega \sum_{n=1}^{\infty}
\frac{B_n P_{2n}(\theta)}{r^{2n+1}}e^{i \Omega t} \right], \\
p_2 & = & {\rm Re} \left[ \, \left( \sum_{n=1}^{\infty} \frac{B_n
P_{2n}(\theta)}{r^{2n+1}}-r\theta \right) e^{i \Omega t} \right], \\
\zeta_2 & = & -{\rm Re} \left[ \, \sum_{n=1}^{\infty} (2n+1)B_n
P_{2n}(\theta) e^{i \Omega t} \right],
\end{eqnarray}
\end{subequations}
\noindent where the expansion coefficients read
\begin{subequations}
\label{sec:WC-cffs2}
\begin{eqnarray}
B_n & = & -\Omega^2\frac{(4n+1) P_{2n}(0)
C_2+\alpha_n}{\Omega_n^2-\Omega^2}
\;\;\; (n>0), \label{sec:WC-cffs-Bn} \\
C_2 & = & \frac{\sum_{n=1}^{\infty} \frac{(2n+1)\alpha_n
P_{2n}(0)}{\Omega_n^2-\Omega^2}}{\Omega^2 \sum_{n=1}^{\infty}
\frac{\alpha_n P_{2n}(0)}{\Omega_n^2-\Omega^2}
-\frac{1}{2}+\frac{i\lambda}{\Omega}}.
\end{eqnarray}
\end{subequations}

The dependence of $\zeta_{20}\equiv \zeta_2(\theta=0)$ on $\Omega$
for solution (\ref{sec:WC-ansatz2}), (\ref{sec:WC-cffs2}) is
plotted in Fig.~\ref{fig8}(b). As before, we observe a significant
increase of the amplitude of the contact line oscillations with
the growth of $\lambda$. Note that at its own antiresonant
frequencies, which are the zeros of the numerator of $C_2$, we
have $\zeta_{20}=0$ irrespective of $\lambda$. In particular,
every standing wave, which is referred by index $n$, has its own
phase shift, i.e., solution (\ref{sec:WC-ansatz2}) with
(\ref{sec:WC-cffs2}) describes traveling waves propagating along
the bubble surface. Although not exactly the same, these phenomena
related to the antiresonant frequencies and traveling waves are
rather similar to those found in
Ref.~\onlinecite{lyubimov-etal-06}.

Having discussed separate contributions in
Eqs.~(\ref{sec:WC-ansatz12}), we can make the final note about the
full solution. The expansion coefficients of this solution in
terms of series in spherical harmonics are a superposition
$h^{-1}B_n+\Pi_0^{-1}A_n$, where both the contributions are small.
The second term becomes negligible compared with the first one
provided that
\begin{equation}
\rho\omega^2 R H \ll P_g. \nonumber
\end{equation}
\noindent This requirement ensures that the compressibility of a
bubble is negligible and incompressible approximation is valid. We
note that a similar inequality has been applied in
Ref.~\onlinecite{vibro-06} in the context of bubbly media.


\section{Conclusions} \label{sec:conclusions}

We have investigated natural and forced oscillations of a
compressible hemispherical bubble put upon a solid substrate. The
contact line motion has been taken into account by applying the
Hocking boundary condition. We have proven that the {\it
linear} shape and volume oscillations demonstrate interaction,
which is the main qualitative result of our study. Having
performed detailed analysis, we have found out two important
features accompanying this interaction. First, we have detected a double resonance, where independent of
the Hocking parameter an unbounded growth of the amplitude occurs.
Second, we have figured out the general condition that can be used
to neglect the bubble compressibility. This requirement turns out
to be stricter than it might be expected. Finally, although our
analysis bases on the assumption of adiabatic oscillations, we
indicate below how the obtained results can be used for an
arbitrary polytropic process in the gas.

We have focused on the natural oscillations and analyzed the
eigenfrequency spectrum. Particularly, we have addressed the
question as to how wetting at the contact line, which is governed
by the Hocking parameter $\lambda$, influences the oscillation
damping. We have neglected viscous and heat dissipation as well as
acoustic irradiation loss. However, because the applied Hocking
condition includes its own energy dissipation mechanism, the
eigenoscillations are generally damped.  The dependence of
eigenfrequencies and the decay rate on $\lambda$ are qualitatively
similar to the case of a drop\cite{lyubimov-etal-06} or an
incompressible bubble.\cite{lyubimov-etal-04} Here, the decay rate
is maximal for $\lambda=O(1)$ and tends to zero at the limits of
the fixed contact line ($\lambda \to 0$) and the fixed contact
angle ($\lambda \to \infty$). In the latter case, the
eigefrequencies of the hemispherical bubble refer directly to
those of the even eigenmodes for a spherical bubble of the same
radius: the breathing mode frequency $\Omega_0^2=\Pi_0^2-2$ and
the frequencies of the shape oscillations
$\Omega_k^2=(2k-1)(2k+1)(2k+2)$, $k>0$.

However, the compressibility of the bubble leads to a number of
peculiar effects. Generally, the eigenfrequency spectrum becomes
dependent on an additional parameter, dimensionless pressure in
the bubble $\Pi_0$. This dependence results in nontrivial
interaction of the volume and shape oscillation and newly found
rearrangement of branches in the spectrum, see Fig.~\ref{fig4}. As
a particular consequence of the rearrangement, the
eigenfrequencies of the compressible bubble are able to not only
monotonically decrease with $\lambda$, as in the case of the drop
or incompressible bubble, but also monotonically grow. Of special
attention is the ``resonant'' case when $\Omega_0\approx\Omega_k$,
characterized by weak dissipation. At the exact coincidence of
these frequencies the contact line is motionless and there is no
dissipation irrespective of $\lambda$. The bubble surface dynamics
corresponds to a superposition of the $k$-th mode of the shape
oscillations and the antiphase radial pulsation.

We have considered normal vibrations of the substrate and studied
the problem of the forced oscillations. In this situation the main
role is played by the spatially uniform pulsations of the pressure
field, which cause the volume oscillations of the bubble. Through
the interaction with the substrate via the moving contact line
these volume oscillations induce the shape oscillations. The
performed analysis of the forced oscillations has shown resonance
phenomena to exist. Particularly, for weak dissipation we have
obtained analytical expressions for the oscillation amplitudes
valid close to the resonance. We have also found out the double
resonance, $\Omega \approx \Omega_0 \approx \Omega_k$, where
$\Omega$ is the frequency of substrate vibration. As it follows
from (\ref{sec:FO-A-res}), in this oscillation regime the resonant
amplitude $A_{res}\propto(\Omega_0^2-\Omega_k^2)^{-2}$. The
divergence of $A_{res}$ as $\Omega\to\Omega_0=\Omega_k$ follows
directly from the problem of natural oscillations, where the
specific case $\Omega_0 = \Omega_k$ predicts no damping of
oscillations.

We indicate that although we have applied the adiabatic law
leading to Eq.~(\ref{sec:NO-bc_pg}), our analysis holds for an
arbitrary polytropic process. In this case, the adiabatic exponent
$\gamma$, which enters Eq.~(\ref{sec:NO-bc_pg}) and the parameter
$\Pi_0$, should be now replaced with the polytropic exponent, $m$.
This generalization makes our theory applicable to a wider range
of bubble sizes and allows for the description of smaller bubbles,
for which heat conductivity becomes the dominant dissipative
effect.\cite{nigmatulin-91,wijngaarden-72} For instance, low
frequency oscillations, $\omega \ll \chi_g R^{-2}$, are governed
by the isothermal process, $m=1$. We note, however, that this
generalization is appropriate only for the forced oscillations.
For natural oscillations, any polytropic process is accompanied by
intensive additional damping, which is different from the Hocking
mechanism. Independent of $\lambda$, this damping results in the
complete attenuation of oscillations within a few periods.

We have considered the special case of weakly compressible bubble
and obtained the criterion identifying whether the bubble
compressibility is insignificant. We have shown that the
compressibility can be neglected only if the dimensionless
pressure $\Pi_0$ in the bubble is large compared to the large $h$,
which is the ratio of the layer depth $H$ to the averaged bubble
radius $R$ (see Fig.~\ref{fig5}). Our analysis allows us to draw a
general conclusion about the impact of compressibility under the
action of vibrations. It might be naively expected that
compressibility effects become negligible at small frequencies
$\omega$ in the sense that $\omega/\omega_c \ll 1$, where
$\omega_c$ is a characteristic frequency of the volume
oscillations. For bubble dynamics,\cite{vibro-06} $\omega_c$ is
the frequency of the breathing mode, $\omega_b$, for homogeneous
fluid media\cite{lyubimov-00} it has the meaning of the acoustic
frequency, $\omega_c \simeq c/H$. However, the correct condition
that does guarantee that the compressibility effects can be
neglected is significantly stricter and can be formulated as
$\left(\omega/\omega_c \right)^2 \ll \mu$. Here $\mu$ is a
small parameter typical for a concrete physical situation. For
instance, as we saw for the bubble dynamics $\mu=h^{-1}\ll 1$,
whereas for thermoacoustic convection $\mu=\beta\Theta \ll 1$,
where $\beta$ is the thermal expansion coefficient and $\Theta$ is
the characteristic temperature difference.

\section{Acknowledgments}

We acknowledge fruitful discussions with D.~V.~Lyubimov and
especially grateful for Ref.~\onlinecite{thanks-DV}. S.S. thanks
DAAD for support; the research was partially supported by CRDF
(Grant No. PE-009-0), Russian Foundation for Basic Research (Grant
No. 04-01-00422-a), and the Foundation ``Perm Hydrodynamics.'' A.S. was
supported by the German Science Foundation (DFG, SPP 1164 ``Nano-
and microfluidics,'' Project No. STR 1021/1).


\end{document}